\documentclass[preprint]{aastex63}
\usepackage{color}

\received{}
\revised{}
\accepted{}
\submitjournal{}
\shorttitle{}
\shortauthors{}
\begin{document}
	\title{Bias of reconstructing the dark energy equation of state from  the Pad$\acute{e}$ cosmography}
	
	\correspondingauthor{Hongwei Yu}
	\email{hwyu@hunnu.edu.cn}
	\correspondingauthor{Puxun Wu}
	\email{pxwu@hunnu.edu.cn}
	\author{Yang Liu}  
	\affiliation{Department of Physics and Synergistic Innovation Center for Quantum Effects and Applications, Hunan Normal University, Changsha, Hunan 410081, China} 
	
	\author{Zhengxiang Li}  
	\affiliation{Department of Astronomy, Beijing Normal University, Beijing 100875, China} 
	
	\author{Hongwei  Yu}
	\affiliation{Department of Physics and Synergistic Innovation Center for Quantum Effects and Applications, Hunan Normal University, Changsha, Hunan 410081, China}
	
	\author{Puxun Wu}
	\affiliation{Department of Physics and Synergistic Innovation Center for Quantum Effects and Applications, Hunan Normal University, Changsha, Hunan 410081, China}
\begin{abstract}
Pad$\acute{e}$ cosmography has been widely used to probe the cosmic evolution and to investigate the nature of dark energy. In this paper, we find that the Pad$\acute{e}$ approximant can describe the cosmic evolution better than the standard cosmography, and if  the luminosity distance $d_L(z)$ described by the Pad$\acute{e}$ approximant is used to reconstruct  the dark energy equation of state $w(z)$, then the  reconstructed $w(z)$ will  approach a constant, {\it i.e.} $1/3$ or $0$, when  the  redshift is very high. This result is general since it is independent of the coefficients in the Pad$\acute{e}$ approximant and the value of the present dimensionless matter density parameter.   This intrinsic  character will bias the $w(z)$ reconstruction  and lead to misconception of the property of dark energy. Therefore,  one  must  exercise caution  in investigating  the property of dark energy from Pad$\acute{e}$ cosmography when the high redshift data, {\it i.e.} $z>2$, are included. 	
\end{abstract}

\section{introduction}\label{sec:intro}
Many observations  including the type Ia supernovae (SNIa) \citep{Riess1998,Perlmutter1999Measurements}, the cosmic microwave background radiation (CMB)~\citep{CMB1, CMB2}, the baryon acoustic  oscillation (BAO)~\citep{BAO}, and so on, have indicated that our universe is undergoing an accelerating expansion. 
To explain this observed mystery, usually, an exotic dark energy component is assumed to exist in our universe. The simplest candidate of dark energy is the cosmological constant $\Lambda$, whose equation of state parameter $w$ ($w\equiv \frac{P_{\mathrm{DE}}}{\rho_{\mathrm{DE}}}$) equals to $-1$, where $P_{\mathrm{DE}}$ and $\rho_{\mathrm{DE}}$ are the pressure and energy density of dark energy, respectively. Although the cosmological constant plus cold dark matter ($\Lambda$CDM) model  is well consistent with most of the observational data, it however suffers from both the  fine tuning problem and the coincidence  problem. 
Furthermore, the value of the Hubble constant ($H_0$) of the $\Lambda$CDM  model determined from the CMB data is in  severe tension with that from the nearby SNIa~\citep{Riess2018,Riess2018M,Planck,Wu}.  
Alternatively, many other dark energy models,  the quintessence scalar field dark energy model~\citep{1988Cosmological}, for instance, have been proposed to explain the current accelerated cosmic expansion. When these dark energy models are used to study the property of dark energy, the  result obtained is unavoidably model-dependent.  
To circumvent this dependence, one may directly parametrize the equation of state $w(z)$ of dark energy in the investigation of the property of dark energy.  One such popular model is the CPL parametrization~\citep{Chevallier2001,Linder2003}. 
Apparently, the results  depend on the parametrization  forms.

A better way to study the property of dark energy is to directly reconstruct it from  cosmological observations. In this    
regard, the  cosmography  is a popular method used to probe the cosmic expansion history and the property of dark energy~\citep{Visser, Dunsby, Capozziello2019b,Bargiacchi,Benetti}.   The standard cosmography (SC) is to Taylor expand the  luminosity distance-redshift  relation $D_L(z)$ at the present time $t_0$ or the present redshift $z_0=0$. Taking a truncation on the Taylor series, one can express $D_L(z)$ as a function of redshift by using some constants 
	\begin{eqnarray}\label{dl-sc}
	D_L(z)&=&\frac{c}{H_0} d_L(z)=\frac{c}{H_0} z\left(1+\alpha_1z+\alpha_2z^2+\alpha_3z^3+\alpha_4z^4+\cdots\right),\end{eqnarray}
where  $d_L(z)$ is called the $H_0$ free luminosity distance which is dimensionless, $c$ is the speed of light,    and  \begin{eqnarray}	
	\alpha_1&=&\frac{1}{2}\left(1-q_0\right),\nonumber\\
	\alpha_2&=&-\frac{1}{6}\left(1-q_0-3q_0^2+j_0\right),\nonumber\\
	\alpha_3&=&\frac{1}{24}\left(2-2q_0-15q_0^2-15q_0^3+5j_0+10q_0j_0+s_0\right),\nonumber\\
	\alpha_4&=&-\frac{1}{20}-\frac{9}{40}j_0+\frac{1}{12} j_0^2-\frac{1}{120}l_0+\frac{1}{20} q_0 -\frac{11}{12}j_0q_0\nonumber\\ && \quad + \frac{27}{40}q_0^2+\frac{7}{8}j_0q_0^2+\frac{11}{8} q_0^3+\frac{7}{8}q_0^4-\frac{11}{120}s_0-\frac{1}{8}q_0s_0 . 
	\end{eqnarray}
	Here, $q_0$, $j_0$, $s_0$ and $l_0$ are the present  \textit{deceleration}, \textit{jerk}, \textit{snap} and \textit{lerk} parameters, respectively. 
 Using observational data to constrain these parameters, one can obtain  $d_L(z)$, and then derive the cosmic evolution and the property of dark energy.  Nevertheless, the convergence region of the Taylor series of the luminosity distance $d_L$ is only around $z=0$, and thus the results from the SC  will be unreliable when the observational data with $z>1$ are used. This  is \textit{the convergence problem} of the cosmographic approach~\citep{Cattoen2007}, and it becomes more and more severe as the observational distance increases.

To alleviate this convergence problem, two different methods are proposed: the $y$-variable method and the Pad$\acute{e}$ one. The former is to enlarge the convergence radius of the cosmographic approach by using a relationship \citep{Cattoen2007,Aviles2012}, {\it e.g.} $y=z/(1+z)$, to replace the redshift variable $z$, so that  $y\to constant$ when $z\to \infty$. The Pad$\acute{e}$ method is based on the rational function approximation \citep{Pade1892}, which can alleviate the convergence problem by exploiting the difference in the order of the numerator and denominator. 
The cosmography based on the  Pad$\acute{e}$ polynomials has been  used widely to discuss the cosmic kinematics and investigate the property of dark energy \citep{Adachi2012,Wei2013,Aviles2014,Gruber2014,Dunsby,Zaninetti2016,Zhou2016,Capozziello2019b,Capozziello2019,Capozziello2020High}. Recent studies have indicated that the Pad$\acute{e}$ approximation has a better fit to high redshift data than the $y$-variable \citep{Capozziello2020High}.
On the other hand, using the Pad$\acute{e}$ approximant for the luminosity distance, the dark energy equation of state  can  be directly reconstructed~\citep{Saini2000,Huterer2001,Sahni2006}.
Direct reconstruction of the dark energy equation of state would involve a  \textit{second-order derivative} of the distance with respect to redshift.  This means that small deviations in  fitting the luminosity distance $d_L(z)$ will be magnified in the reconstruction of $w(z)$, and will cause unstable results~\citep{Huterer2001,Maor2001,Gerke2002}.
In this paper,  we find that there is indeed an intrinsic defect in deriving the dark energy equation of state from the Pad$\acute {e}$ approximation of the luminosity distance. When $z$ is very large, $w(z)$ given from the luminosity distance $d_L(z)$ described by the Pad$\acute{e}$ polynomial  approaches  a constant, {\it i.e.} $0$ or $1/3$, which may result in a bias for the property of dark energy reconstructed from observational data with the Pad$\acute{e}$ approximant. Therefore, although the Pad$\acute{e}$ cosmography is a viable way to investigate the cosmic evolution, it may lead to some unreliable results on  the property of dark energy.  

 This paper is arranged as follows. In Section \ref{pade}, we briefly review how to approximate the luminosity distance using the Pad$\acute{e}$ approximation, and diagnose the Pad$\acute{e}$ cosmography by using the one-parameter method. We analyze the evolution of the reconstructed $w(z)$ from the Pad$\acute{e}$ approximant at high redshifts in Section \ref{fit}. Finally, we conclude in Section \ref{conclusion}.
 
\section{the Pad$\acute{e}$ cosmography}\label{pade}

\subsection{Pad$\acute{e}$ approximant}

The Pad$\acute{e}$ approximant is obtained by expanding a function as a ration of numerator and denominator power series. Its radii of convergence are usually broader than that of a Taylor series since the  Taylor expansion converges only near the expansion point. For a generic function $f(x)$, its Taylor series expansion has the form  $f(x)=\sum_{i=0}^{\infty}c_i x^i$, which converges in some neighborhood of the origin.  The Pad$\acute{e}$ approximant of order $(n,m)$ to $f(x)$ is
defined to be a rational function $P_{n,m}(x)$ expressed in a fractional form:   
\begin{equation}\label{2}
P_{n,m}(x)=\frac{\sum_{i=0}^{n}a_ix^i}{1+\sum_{i=1}^{m}b_i x^i},
\end{equation}
where $n$ and $m$ denote the highest order of the \textit{numerator} and \textit{denominator}, respectively. The total order of the Pad$\acute{e}$ polynomial is $n+m$.  Approximating this function at $x=0$, the coefficients ($a_0,\dots,a_n,b_1\dots,b_m$) can be obtained in the following way:
\begin{eqnarray}\label{3}
P_{n,m}(0)&=f(0) , \nonumber\\
P'_{n,m}(0)&=f'(0) , \nonumber\\
\dots & \nonumber\\
P^{(n+m)}_{n,m}(0)&=f^{(n+m)}(0) ,
\end{eqnarray}
where $P'_{n,m}(0)$ is the first-order derivative of $P_{n,m}(x)$ at $x=0$ and $P^{(n+m)}_{n,m}(0)$ is the $n+m$ order derivative of $P_{n,m}(x)$ at $x=0$.  

The coefficients of the Pad$\acute{e}$ approximant can also be obtained as follows.  If the Taylor series of  $f(x)$ is   truncated at order $k$, $f(x)=\sum_{i=0}^{k}c_i x^i$, one can assume:
\begin{equation}\label{4}
c_0+c_1 x+c_2 x^2+\dots+c_k x^k=\frac{a_0+a_1 x+\dots+a_n x^n}{1+b_1 x+\dots+b_m x^m},
\end{equation}
where $k$ must satisfy the relation $k\geq n+m$, otherwise the whole set of linear equations will not be obtained. Multiplying both sides of this equation by the denominator ($1+b_1 x+\dots+b_m x^m$), and setting coefficients of the same order to be equal, we have
\begin{eqnarray}\label{pade-relation}
&c_0=a_0 , \nonumber\\
&c_1+c_0b_1=a_1 ,\nonumber\\
&\dots\nonumber\\
&c_n+c_{n-1}b_1+\dots+c_0b_{n}=a_n ,\nonumber\\
&c_{n+1}+c_nb_1+\dots+c_0b_{n+1}=0 ,\\
&\dots \nonumber.
\end{eqnarray}
By solving this system of linear  equations, the value of coefficients ($a_i,b_i$) can be obtained. It is easy to see that  a function, which can be expanded in a Taylor series,  can always be approximated by a Pad$\acute{e}$ approximant.


\subsection{ One-parameter diagnostic on Pad$\acute{e}$ cosmography }

 The Pad$\acute{e}$ approximation has been widely used in cosmology \citep{Wei2013,Gruber2014,Mehrabi2018,Capozziello2019,Rezaei2019,Rezaei2021}.  
Usually, we can use the Pad$\acute{e}$ approximant to express the luminosity distance $d_L(z)$: 
\begin{equation}\label{dl-pade}
	d_L(z)=P_{n,m}(z)=\frac{z+\sum_{i=2}^{n}a_iz^i}{1+\sum_{i=1}^{m}b_i z^i},
\end{equation}
where   $a_0=0$ and $a_1=1$ have been used,  which arise from the requirements that $d_L(z)=0$ and $d'_L(z)=1$ must be satisfied at $z=0$.  Since $d_L(z)$ is an increasing function of redshift, it is required that $n\geq m$ in Eq.~(\ref{dl-pade}).   

The discussion in the above subsection shows that one can use the Pad$\acute{e}$ approximant to   approximate the SC. In this case,  the coefficients ($a_i,b_i$) in the Pad$\acute{e}$ approximants can be expressed with the cosmological parameters ($q_0,j_0,{\it etc}$) by solving  Eq.~(\ref{pade-relation}). 
In the framework of  the flat $\Lambda$CDM model, the Hubble parameter has the form  
\begin{equation}\label{H(z)}
H_{\mathrm{\Lambda CDM}}(z)=H_0 \sqrt{\Omega_{\mathrm m0}(1+z)^3+1-\Omega_{\mathrm m0}},
\end{equation}
after neglecting the  radiation energy density, where  $\Omega_{\mathrm m0}$ is the present density parameter of pressureless matter. Then,   the  cosmological parameters $q_0,j_0, s_0$ and  $l_0$ can be, respectively,  expressed as
	\begin{eqnarray}
		q_0&=&-1+\frac{3}{2}\Omega_{\mathrm{m0}}, \nonumber\\
		j_0&=&1, \nonumber\\
		s_0&=&1-\frac{9}{2}\Omega_{\mathrm{m0}}, \nonumber\\
		l_0&=&1+3\Omega_{\mathrm{m0}}-\frac{27}{2}\Omega_{\mathrm{m0}}^2.
	\end{eqnarray}
The values  of these parameters depend only on $\Omega_{\mathrm{m0}}$, which also determines the coefficients ($a_i, b_j$) when using the Pad$\acute{e}$ approximant to approximate  the SC.  Using data to constrain $\Omega_{\mathrm{m0}}$, we can discuss the deviation of the cosmography from the $\Lambda$CDM. This method is called the one-parameter diagnostic, which has been used to investigate the SC in  \citep{Aviles 2017}. 

To diagnose the Pad$\acute{e}$ cosmography with the one-parameter method, 
we use the Pantheon SNIa sample~\citep{Scolnic:2017caz}, which consists of 1024 data points. We consider four popular Pad$\acute{e}$ approximants to express  the luminosity distance $d_L(z)$~\citep{Aviles2014,Capozziello2020High}: 
\begin{eqnarray}\label{four pade}
	P_{2,1}&=&\frac{z+a_2z^2}{1+b_1z}, \nonumber\\
	P_{2,2}&=&\frac{z+a_2z^2}{1+b_1z+b_2z^2},\nonumber\\
	P_{3,1}&=&\frac{z+a_2z^2+a_3z^3}{1+b_1z},\nonumber\\
	P_{3,2}&=&\frac{z+a_2z^2+a_3z^3}{1+b_1z+b_2z^2}.
\end{eqnarray}
It has been found that these Pad$\acute{e}$ approximants have small deviations  from  the $\Lambda$CDM model~\citep{Aviles2014,Capozziello2020High} by using  different  observational data including SNIa~\citep{Suzuki:2011,Scolnic:2017caz}, BAO~\citep{Percival:2009xn}, the Observational Hubble Data~\citep{Jimenez:2001gg} and the CMB shift parameter~\citep{Ade2015} to constrain the model parameters. 

As a comparison, we also diagnose  the SC with the one-parameter method. We use SC$_n$ to denote the n-th order luminosity distance. For example, SC$_3$ represents the 3rd order luminosity distance.  Constraints on $\Omega_{\mathrm{m0}}$  can be obtained by minimizing the $\chi$-square:
\begin{equation}\label{chi2}
\chi^2=\sum_{i=1}^{N}\left[\frac{\mu^{obs}_i(z_i)-\mu^{th}_i(H_0, \Omega_{\mathrm{m0}}, z_i)}
{\sigma^{obs}_{\mu_i}}\right]^2,
\end{equation}
where $N=1048$, $\mu^{obs}_i$ and $\sigma^{obs}_{\mu_i}$ are the distance modulus and the corresponding error of the Pantheon SNIa, respectively,
and $\mu^{th}_i=25+5\log[D_L(z_i)/\mathrm{Mpc}]$ is the distance modulus from the SC or the Pad$\acute{e}$ approximant at $z_i$.
In Eq.~(\ref{chi2}), $H_0$ is a noise parameter, which is marginalized by using the method given in \citep{Nesseris2004}.  In our analysis,   the  {\it CosmoMC} code is used \footnote{The {\it CosmoMC} code is available at \href{https://cosmologist.info/cosmomc/}{https://cosmologist.info/cosmomc}.}.

\begin{figure}
	\includegraphics[width=0.5\textwidth]{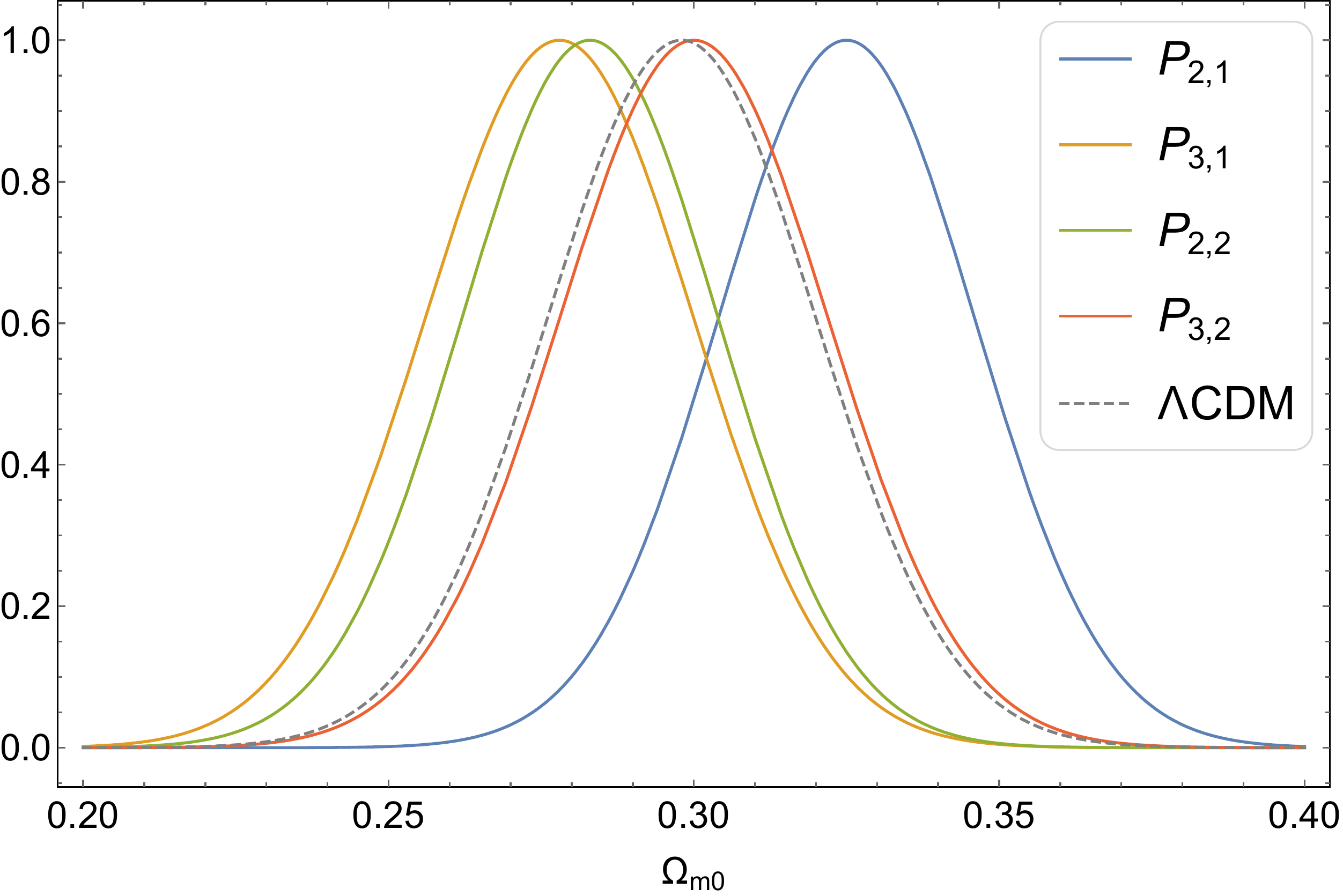}
	\includegraphics[width=0.5\textwidth]{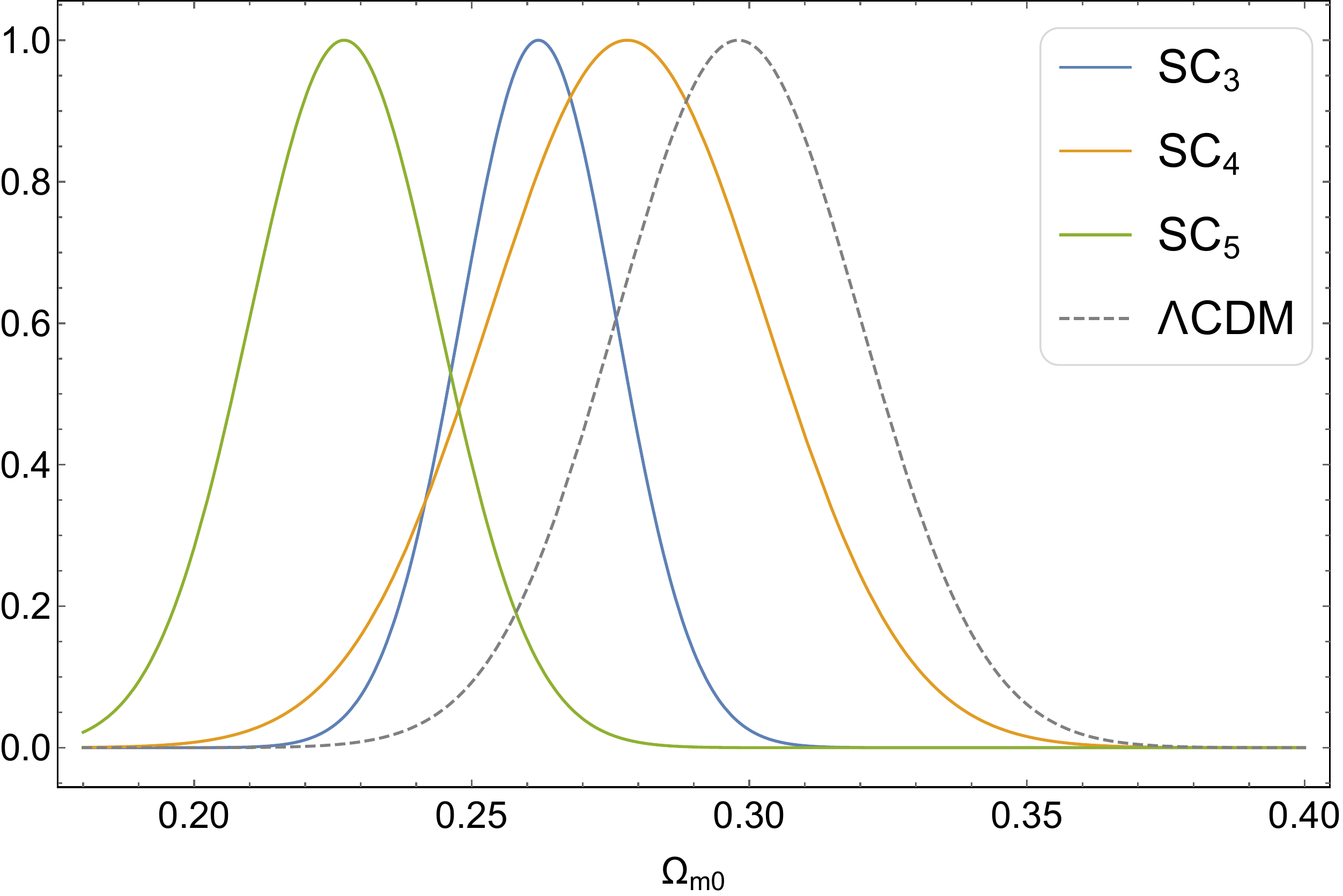}
	\caption{ 
		One-parameter diagnostic on the Pad$\acute{e}$ cosmography (left panel) and the standard one (right panel). 
\label{one-par}}
\end{figure}

The results are shown in Figure \ref{one-par} and the best fit values of $\Omega_{\mathrm{m0}}$ at the $1\sigma$ confidence level ({\it CL}), respectively,  are:
	\begin{eqnarray}
		&&\Omega_{\mathrm{m0}}|_{P_{2,1}}=0.325\pm 0.021,\nonumber\\
		&&\Omega_{\mathrm{m0}}|_{P_{3,1}}=0.278\pm 0.022,\nonumber\\
		&&\Omega_{\mathrm{m0}}|_{P_{2,2}}=0.283\pm 0.021,\nonumber\\
		&&\Omega_{\mathrm{m0}}|_{P_{3,2}}=0.300\pm 0.022,\nonumber\\
		&&\Omega_{\mathrm{m0}}|_{SC_{3}}=0.262\pm 0.014,\nonumber\\
		&&\Omega_{\mathrm{m0}}|_{SC_{4}}=0.278\pm 0.025,\nonumber\\
		&&\Omega_{\mathrm{m0}}|_{SC_{5}}=0.227\pm 0.017,\nonumber\\
		&&\Omega_{\mathrm{m0}}|_{\Lambda CDM}=0.298\pm 0.022.
	\end{eqnarray}
It is easy to see that when the number of model parameters  is the same, the one-parameter diagnostic shows that the Pad$\acute{e}$ cosmography is apparently better than the SC. Although $P_{2,1}$, $P_{3,1}$ and $P_{2,2}$ are consistent with the $\Lambda$CDM only at the margin of $1\sigma$,    $P_{3,2}$ gives  almost the same result as the $\Lambda$CDM.   Thus, the Pad$\acute{e}$ cosmography can provide an unbiased estimation of the background cosmology. 
 
\begin{figure}
	
	\gridline{
		\fig{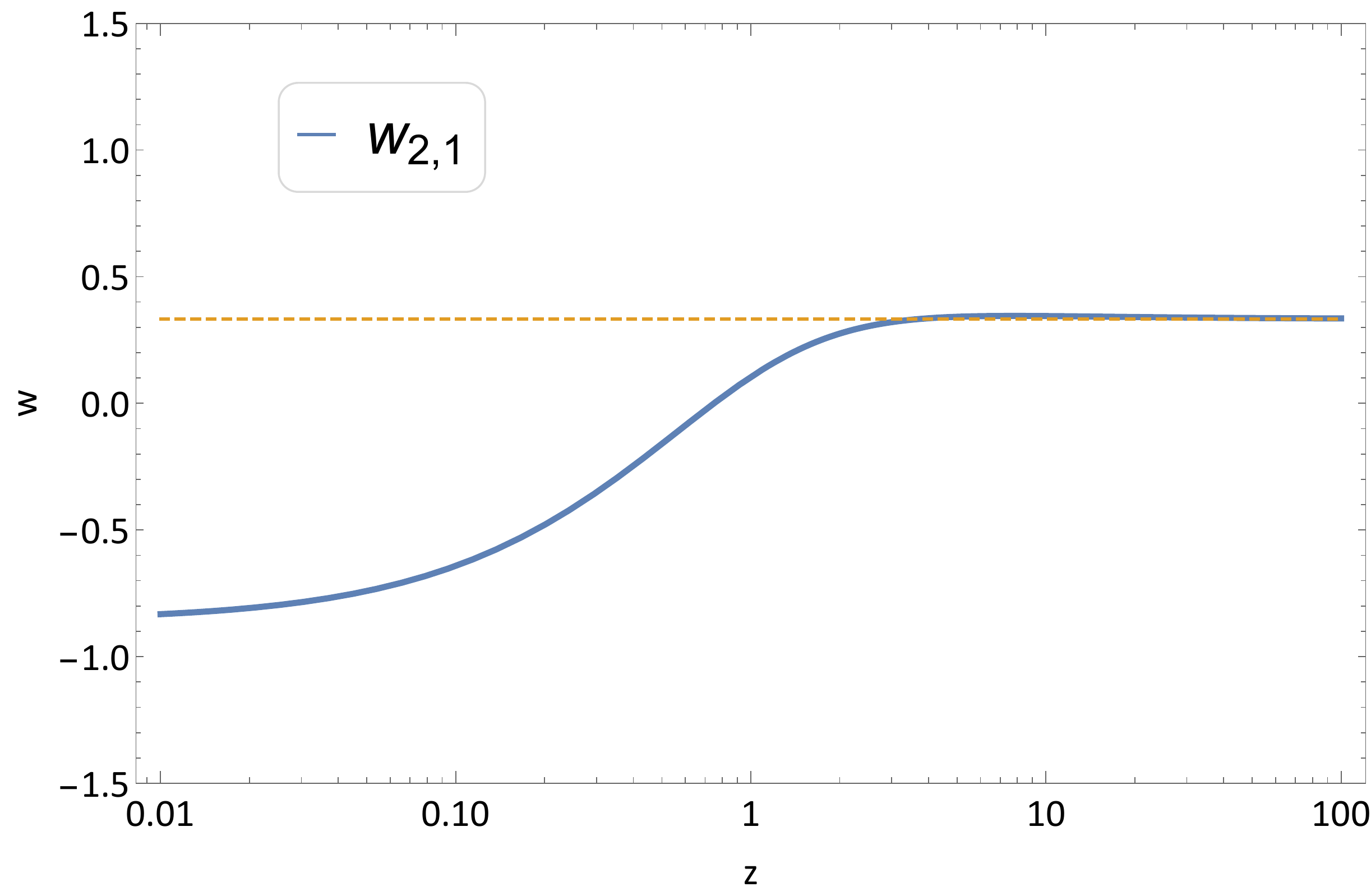}{0.5\textwidth}{(a)}
		\fig{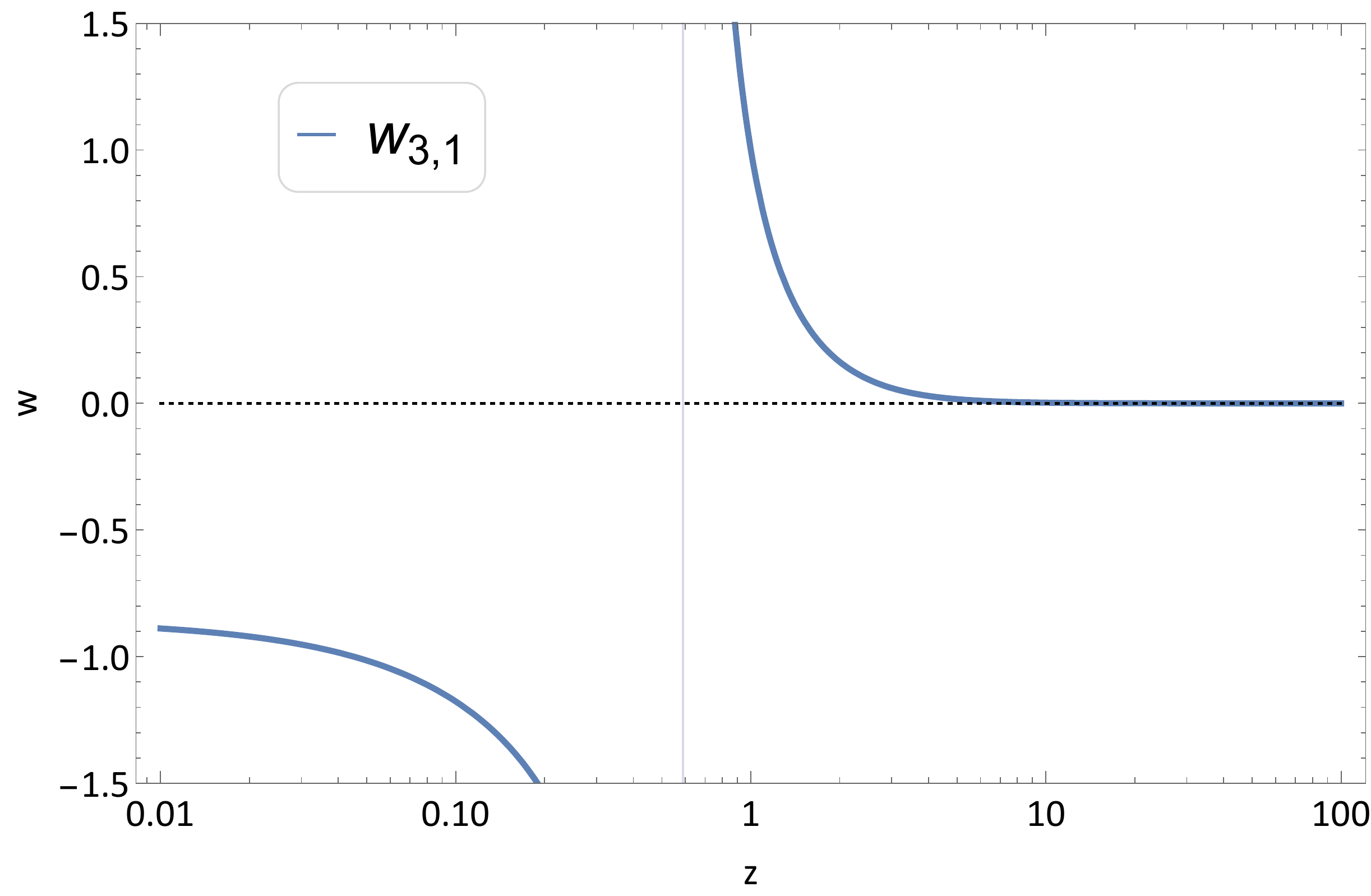}{0.5\textwidth}{(b)}
	}
	\gridline{
		\fig{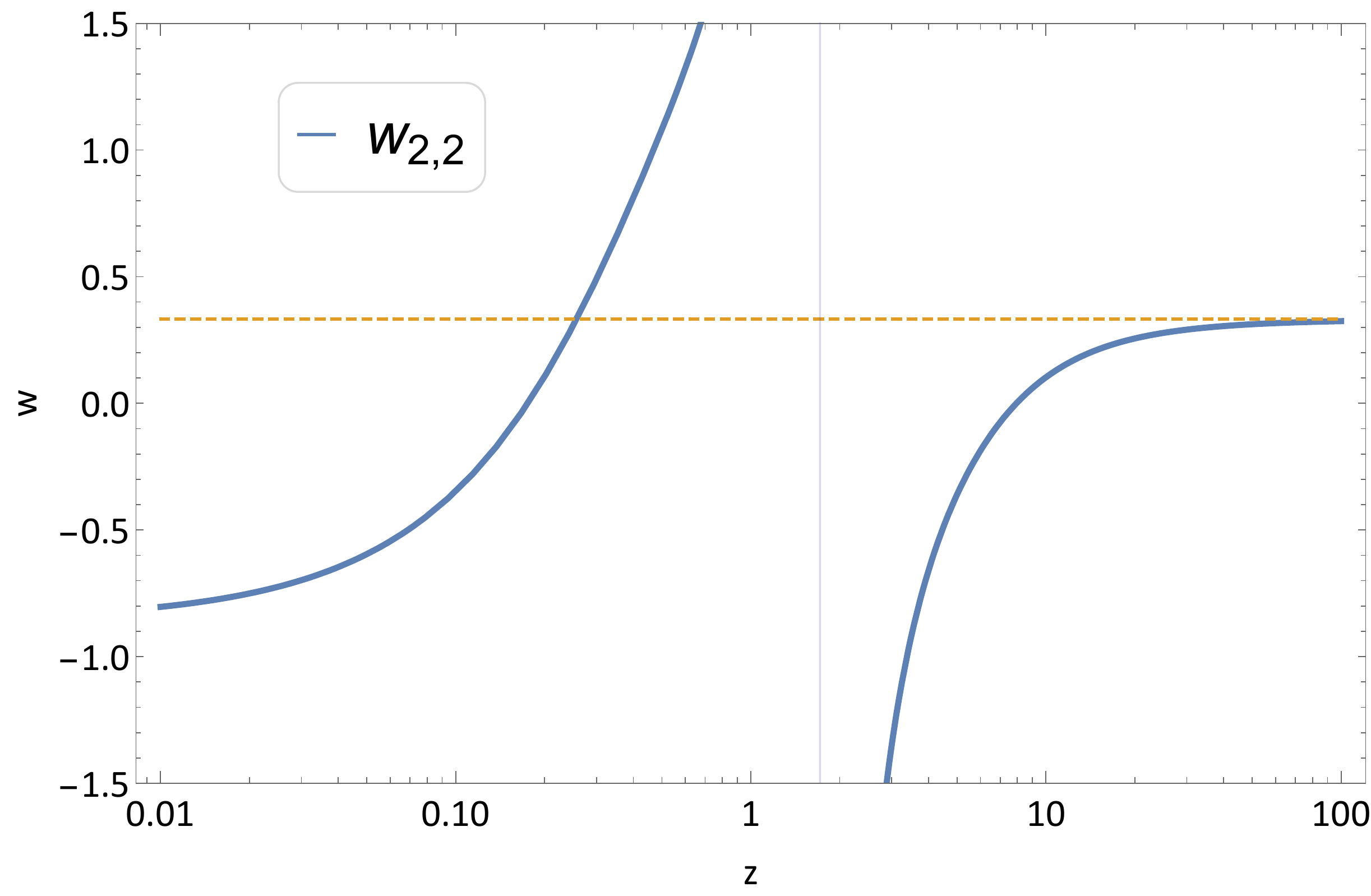}{0.5\textwidth}{(c)}
		\fig{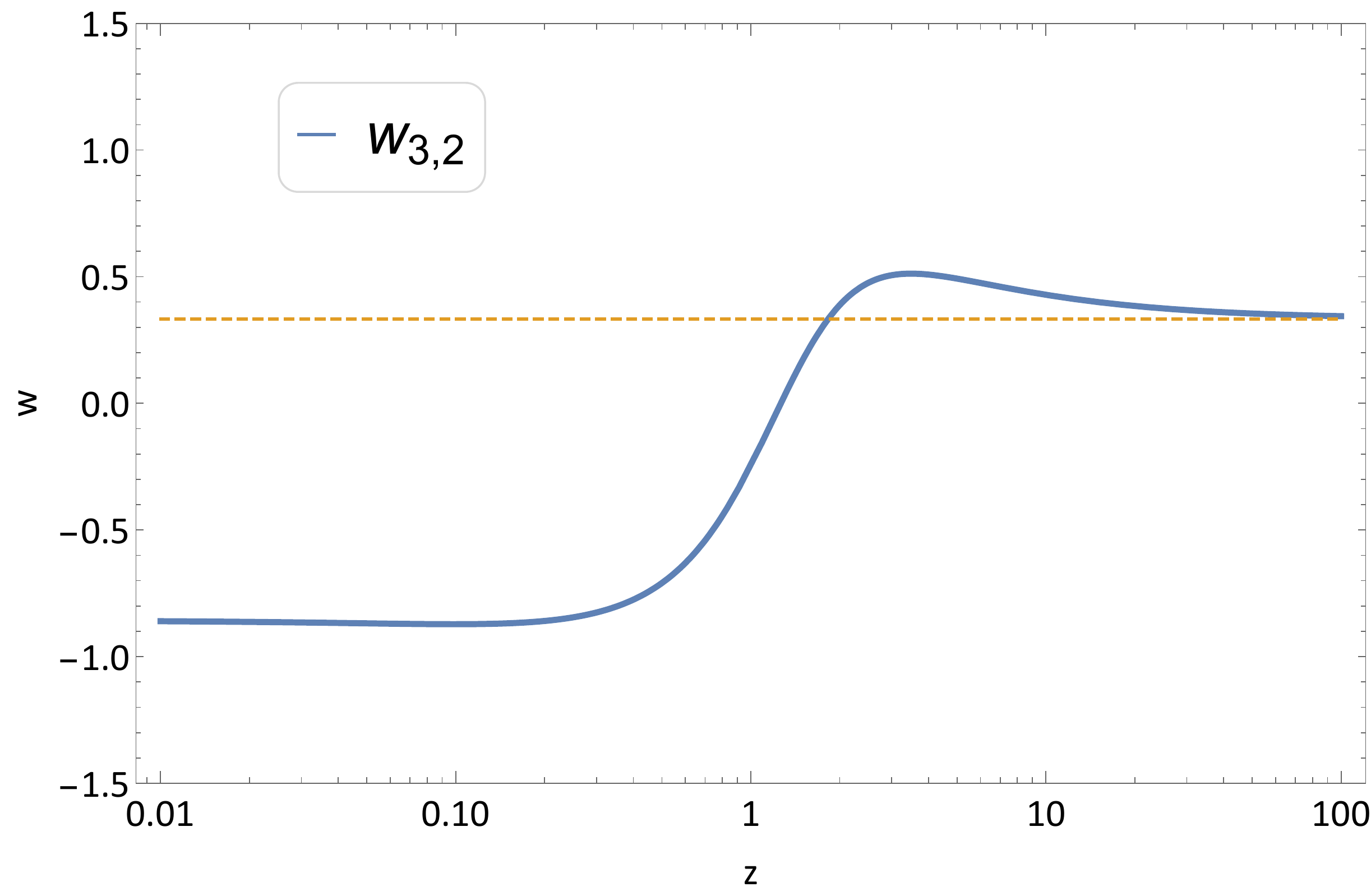}{0.5\textwidth}{(d)}
	}
	\caption{
		 The evolution of the reconstructed dark energy equation of state from the Pad$\acute{e}$ cosmography.  $w_{2,1}$, $w_{3,1}$, $w_{2,2}$ and $w_{3,2}$ represent the results from  $P_{2,1}$, $P_{3,1}$, $P_{2,2}$ and $P_{3,2}$, respectively. 
		The coefficients in the Pad$\acute{e}$ approximants are set to be $a_2=1.7$, $a_3=1$, $b_1=1$, and $b_2=0.5$, and   $\Omega_{\mathrm{m0}}$ is set to be $0.3$. The black dotted  and orange dashed lines denote $w=0$ and $w=1/3$, respectively.  }
		\label{fig_ln_w}
	
\end{figure}

{\section{Reconstructing  the equation of state of dark energy from Pad$\acute{e}$ cosmography}\label{fit}
\subsection{$w(z)$ reconstructed from  Pad$\acute{e}$ approximant}
 
Using the Pad$\acute{e}$ approximant to approximate  the luminosity distance, one can obtain the expression of the Hubble parameter 
\begin{equation}\label{h with pade}
H(z)=\frac{(z+1)^2 \left[\sum _{i=1}^m (b_i z^i+1)\right]^2}{(z+1) \left[\sum _{i=0}^n i a_i z^{i-1} \sum_{i=1}^m \left(b_i z^i+1\right)- \sum _{i=0}^n a_i z^i  \sum _{i=1}^m i b_i z^{i-1}\right]-\sum _{i=0}^n
	a_i z^i \sum _{i=1}^m \left(b_i z^i+1\right)}
\end{equation}
after assuming a spatially flat FLRW universe, where Eq.~(\ref{dl-pade}) and the following relation 
 \begin{equation}\label{h}
H(z)=\left[\frac{d}{dz}\left(\frac{d_L(z)}{1+z}\right)\right]^{-1}
\end{equation} 
have been used. 

Assuming that the universe is filled with  pressureless matter  and dark energy, we can obtain  the equation of state of dark energy from the Hubble parameter \begin{equation}\label{w}
w(z)=\frac{(2 (1+z)/3)(\ln H)'-1}{1-(H_0/H)^2\Omega_{\mathrm m0}(1+z)^3}.
\end{equation}
Substituting $H(z)$ (Eq.~(\ref{h with pade})) and its first derivative into Eq.~(\ref{w}), one can  reconstruct the dark energy  equation of state from the Pad$\acute{e}$ approximant. Apparently, $w(z)$ depends on the second derivative of $d_L(z)$ with respect to $z$, which results in that the reconstructed result may be unreliable.  Here, we find that the reconstructed $w(z)$ approaches a constant when $z$ is very large. This will  render the reconstructed result biased.    Since the value of a rational function as $z\to \infty$ is only dependent on the highest-order term in the numerator and denominator, we replace the summation terms, {\it i.e.} $\sum_{i=0}^{n}a_iz^i$ and $\sum_{i=1}^{m}b_iz^i$,  in $w(z)$ with the highest order  term $a_nz^n$ and $b_mz^m$, and then  obtain 
\begin{equation}\label{w inf}
w(z)|_{z\rightarrow \infty}=\frac{A_{0}+A_{-1}z^{-1}}{B_{0}+B_{-1}z^{-1}+B_{2n-2m-1}z^{2n-2m-1}} , 
\end{equation}
where the coefficients are defined as:
\begin{eqnarray}\label{14}
A_{0}\qquad\;\;\:&=&b_m^5 \left(2 m^2+m (3-4 n)+2 n^2-3 n+1\right),\nonumber\\
A_{-1}\qquad\;&=&b_m^5 \left(6 m^2+m (8-12 n)+6 n^2-8 n+1\right),\nonumber\\
B_{0}\qquad\;\;\:&=&3 b_m^5 (m-n+1),\nonumber\\
B_{-1}\qquad\;&=&3 b_m^5 (2 m-2 n+1),\nonumber\\
B_{2n-2m-1}&=&-3 \Omega_{\mathrm{m0}} a_n^2 b_m^3 (m-n+1)^3.
\end{eqnarray}
It is worth noting that when $n=m+1$, the coefficients become $A_0=B_0=B_{2n-2m-1}=0$ and $A_{-1}=\frac{1}{3}B_{-1}=1$, which implies that $w$ approaches  $1/3$. 
When $n=m$, we have $A_0=\frac{1}{3}B_0=-1$ and thus $w\rightarrow \frac{1}{3}$.
The $n>m+1$ case is different from the previous two cases since the term containing  $B_{2n-2m-1}$ in Eq.~(\ref{w inf}) must be taken into account. Due to $2n-2m-1>0$, it is easy to obtain that  $w(z)\to \frac{A_{0}}{B_{2n-2m-1}z^{2n-2m-1}} \to 0$ when $z$ is very large. 
Therefore, we find that in the high redshift regions the equation of state of dark energy reconstructed from the Pad$\acute{e}$ cosmography will approach  a constant, {\it i.e.} $0$ or $1/3$.  This character will bias our understanding of the property of dark energy.  
 To show this character clearly, we plot in  
Fig.~(\ref{fig_ln_w})  the evolutionary curves of the reconstructed $w(z)$ from four popular Pad$\acute{e}$ approximants  given in Eq.~(\ref{four pade}). In this figure,  $w_{2,1},w_{3,1},w_{2,2}$ and $w_{3,2}$   represent  the dark energy equation of state reconstructed from $P_{2,1},P_{3,1},P_{2,2}$ and $P_{3,2}$, respectively. It is easy to see that when the redshift  $z$  is about $10$, then $w_{2,1}$, $w_{2,2}$ and $w_{3,2}$ converge to $1/3$   and $w_{3,1}$  approaches  0. There exists a singularity in the reconstructed $w_{2,1}$ and $w_{2,2}$,  and the redshift  where  the singularity appears  depends on the values of the coefficients of the Pad$\acute{e}$ approximants.

To further discuss the bias in reconstructing the property of dark energy from the Pad$\acute{e}$ cosmography, in the following we will use the mock SNIa data based on the $\Lambda$CDM model and the real data from the Pantheon SNIa sample~\citep{Scolnic:2017caz}  to constrain the free parameters in the Pad$\acute{e}$ approximants, respectively, and then study the evolution of $w(z)$ from four popular approximants given in Eq.~(\ref{four pade}).   
}

\subsection{Mock data}\label{sec:mock & analys}
Since the maximum redshift of the Pantheon sample is $2.26$ and the total data number is only $1048$ in the one-parameter diagnostic,
the observational data may not give tight constraints on high redshift regions of $w(z)$, and thus are not  effective in analyzing  the bias discussed in the above subsection. Therefore, here we  mock SNIa  distances on the basis of the upcoming \textit{Wide Field InfraRed Survey Telescope (WFIRST)}, which was the highest-ranked large space-based mission of the 2010 decadal surveys. A primary objective of this mission is to precisely constrain the nature of dark energy with multiple probes, including SNIa. Optimistically, the Imaging: Allz strategy of \textit{WFIRST} could collect $\sim13500$ SNIa in the redshift range $0<z<3$ \citep{Hounsell2018}. The number of SNIa that can be discovered by \textit{WFIRST} is expected to follow the volumetric rates \citep{Rodney2014,Graur2014},
\begin{eqnarray}
R_{\mathrm{SNIa}}(z)=\left\{
\begin{array}{rcl}
2.5\times(1+z)^{1.5} (10^{-5}~\mathrm{yr^{-1}Mpc^{-3}}),&z<1, \\
9.7\times(1+z)^{0.5} (10^{-5}~\mathrm{yr^{-1}Mpc^{-3}}),&~~~~~1<z<3 .
\end{array} \right.
\label{equa:rz}
\end{eqnarray}
As the expected detection rate is low for $z>3$ SNIa, we do not simulate events at those redshifts.  For a sample of SNIa, the total uncertainty of the distances $\sigma_\mu$  consists of two main components, the statistical uncertainty $\sigma_{\mu,\mathrm{stat}}$ and the systematic uncertainty $\sigma_{\mu,\mathrm{sys}}$, i.e. $\sigma_{\mu,\rm tot}^2=\sigma_{\mu,\rm stat}^2+\sigma_{\mu,\rm sys}^2$.    With the fractional statistical uncertainty (panel $h$ of Figure 7 in \citep{Hounsell2018}) and all ingredients of  the systematic uncertainty (Figures 9 and 10 in \citep{Hounsell2018}, including wavelength dependent calibration, nonlinearity, contamination, host-mass evolution, intrinsic scatter, population drift, and zero-point uncertainties) taken into consideration, we can generate a simulated distance modulus sample of SNIa in a given fiducial model. The fiducial value $\mu^{\rm fid}(z)$ is obtained from the spatially flat $\Lambda$CDM model with the model parameters being $H_0=70$ $\mathrm{km}$ $\mathrm{s}^{-1}$ $\mathrm{Mpc}^{-1}$ and $\Omega_{\mathrm {m0}}=0.3$.  For a sample containing $N_{\rm SN}=13570$ mocked SNIa in the range $0<z<3$, we draw a random number from the Gaussian distribution $N(\sigma_{\mu,\rm{tot}},\sigma_{\mu,\rm{tot}}/\sqrt{N_{\rm SN}})$ as the uncertainty $\sigma_{\mu}(z)$. 
For simplicity and approximate predictions, we only take the diagonal covariance matrix into account in our analysis. For the distance modulus of the mocked data, we use  $\mu^{\rm mock}(z)=\mu^{\rm fid}(z)$ at each data point in order to compare with the exact $-1$ line.   From this mocked SNIa sample, the coefficients in the Pad$\acute{e}$ approximants can be constrained by minimizing the $\chi$-square given in Eq.~(\ref{chi2}).

 \subsection{results} 
  
\begin{figure}
			\includegraphics[width=0.4\textwidth]{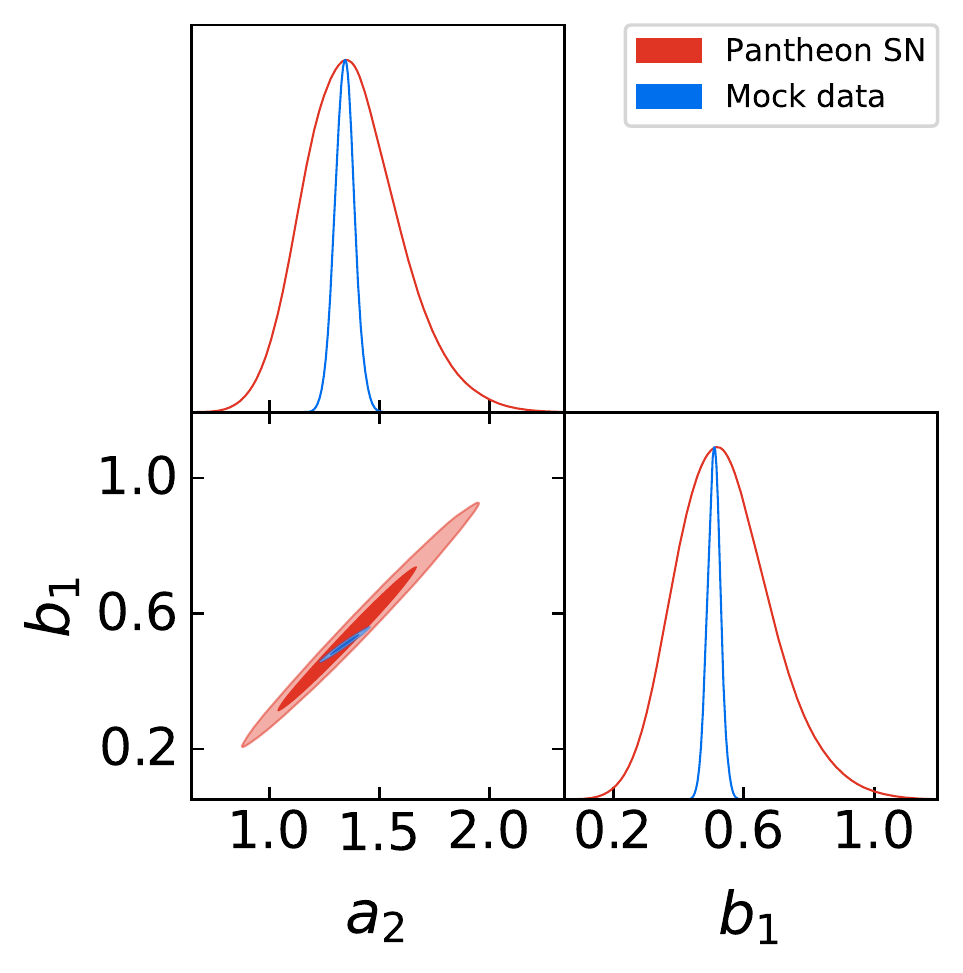}
			\includegraphics[width=0.45\textwidth]{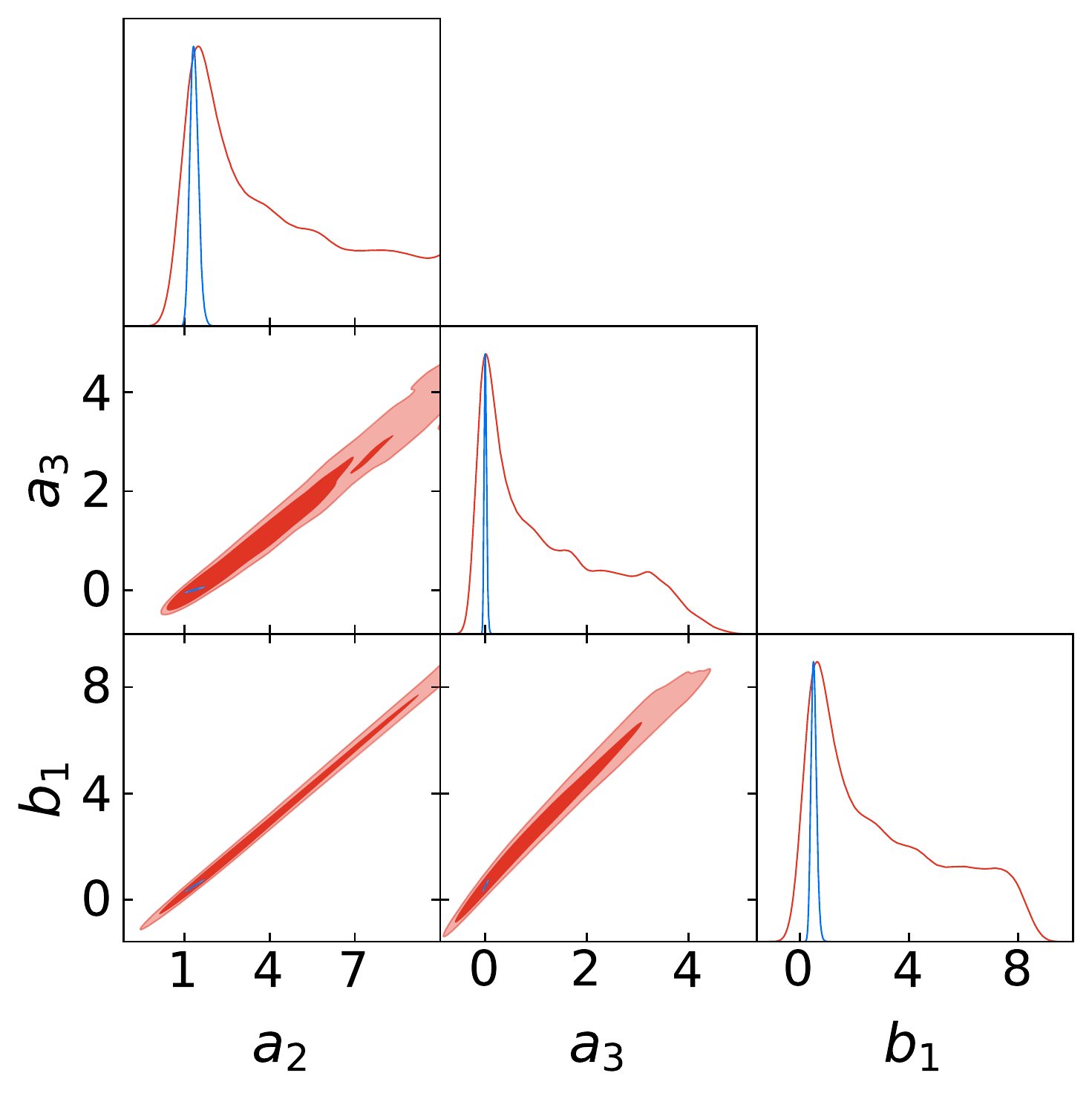}
		\includegraphics[width=0.45\textwidth]{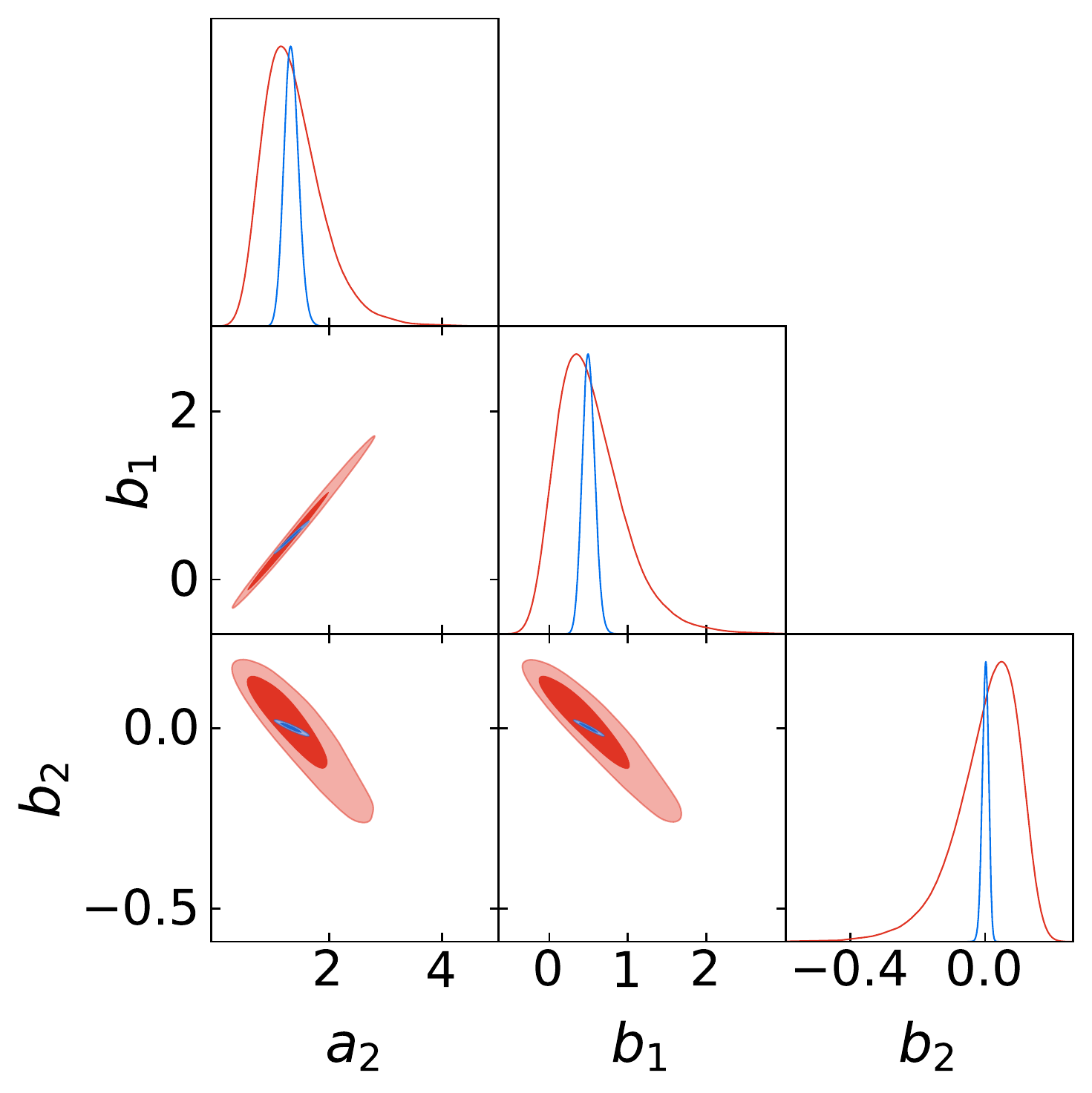}
		\includegraphics[width=0.55\textwidth]{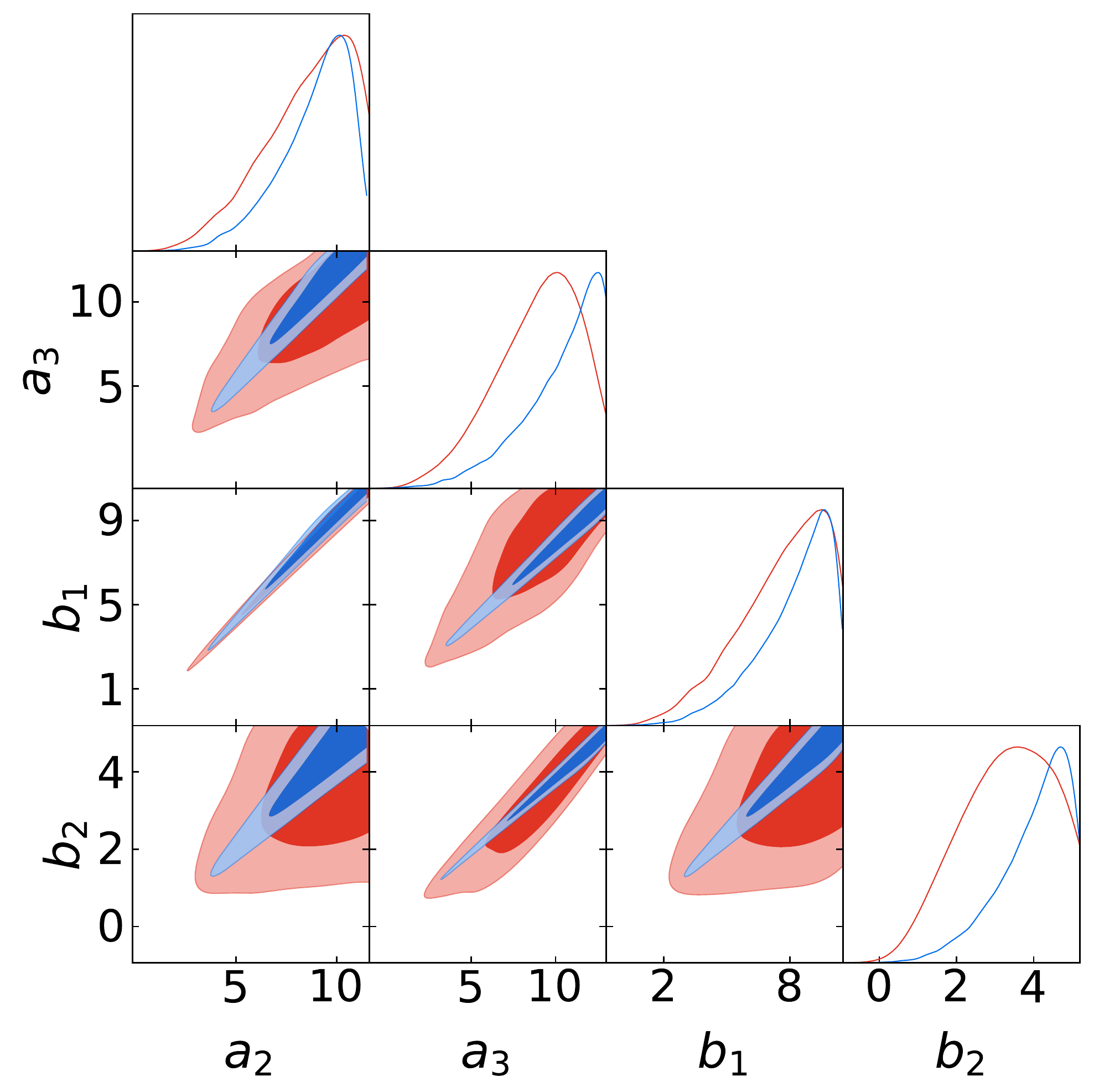}
	\caption{ Contour plots for the coefficients of    four Pad$\acute{e}$ approximants. Top left, top right, bottom left and bottom right represent  $P_{2,1},P_{3,1},P_{2,2}$ and $P_{3,2}$, respectively. The red and blue colors show the constraints from  the Pantheon SNIa sample and the mock data, respectively.
		\label{fig2}}
\end{figure}

 The constraints on coefficients are shown in  Fig.~\ref{fig2} and summarized in Tabs.~\ref{tab1} and \ref{tab2}. Figure~\ref{fig2} gives the contours of the model parameters of the Pad$\acute{e}$ approximants, in which red and blue colors represent the  Pantheon and the mock data, respectively. It is easy to see that the results from the Pantheon sample are consistent with those from the mock data, but the latter gives  very strong constraints.
   
\begin{figure}
 	\centering
 	\includegraphics[width=0.55\textwidth]{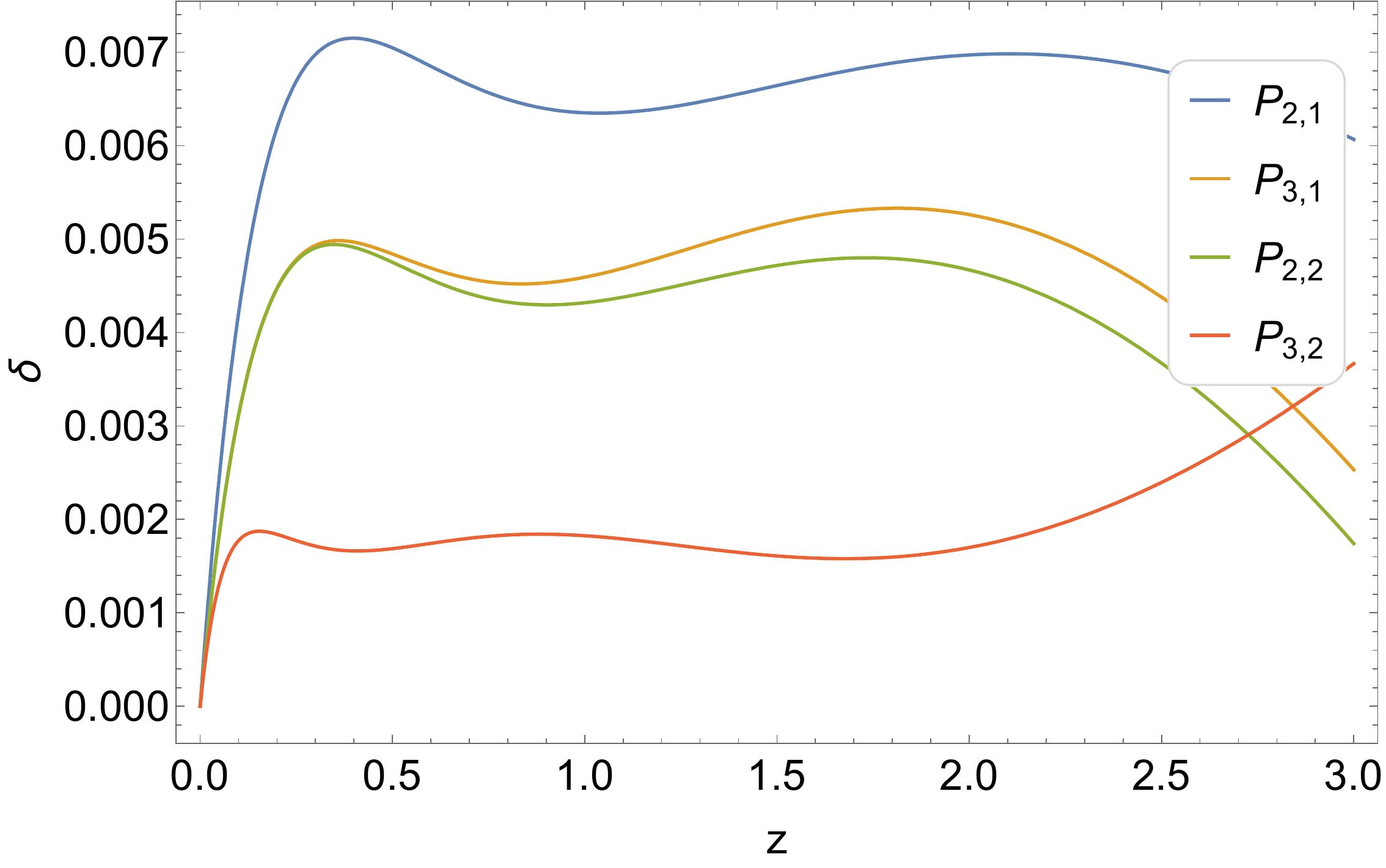}
	\caption{The evolution of the relative deviation $\delta(z)$ obtained from 13570 mock data. 
		\label{fig3}}
\end{figure}

To discuss the deviations of the Pad$\acute{e}$ cosmography from the fiducial model used to mock data, we investigate the evolution of $\delta(z)$, which is defined as:
\begin{equation}\label{13}
\delta(z)=\frac{|d_{L}^{Pad\acute{e}}(z)-d_{L}^{\Lambda CDM}(z)|}{d_{L}^{\Lambda CDM}(z)}.
\end{equation}
In Fig.~\ref{fig3}, we show our results. One can see that four Pad$\acute{e}$ polynomials can describe very well the cosmic evolution since the relative deviation $\delta$ is less than $0.8\%$ in the redshift region between $0$ to $3$, and  $P_{3,2}$ gives the best fit result due to that   $\delta$ is $<0.4\%$ in this case.

\begin{deluxetable}{cccccccccccc}
	\tablenum{1}
	\tablecaption{\label{tab1}}
	\tablewidth{0pt}
	\tablehead{
	& \multicolumn{2}{c}{$P_{2,1}$} & & \multicolumn{2}{c}{$P_{3,1}$} & & \multicolumn{2}{c}{$P_{2,2}$} & & \multicolumn{2}{c}{$P_{3,2}$} \\
	\cline{1-3} \cline{5-6} \cline{8-9} \cline{11-12}
	\colhead{} &\colhead{\it Mean($\sigma$)} &\colhead{\it 0.68 CL} &\colhead{} &\colhead{\it Mean($\sigma$)} &\colhead{\it 0.68 CL} &\colhead{} &\colhead{\it Mean($\sigma$)} &\colhead{\it 0.68 CL} &\colhead{} &\colhead{\it Mean($\sigma$)} &\colhead{\it 0.68 CL}
	}
	\startdata
	$a_2$ & $1.343(0.045)$ & ${}^{+0.045}_{-0.045}$ & & $1.348(0.147)$ & ${}^{+0.127}_{-0.160}$ & & $1.331(0.127)$ & ${}^{+0.118}_{-0.132}$ & & $8.911(1.716)$ & ${}^{+2.205}_{-0.803}$ \\
	$a_3$ & -- & -- & & $0.001(0.025)$ & ${}^{+0.021}_{-0.028}$ & & -- & -- & & $10.483(2.267)$ & ${}^{+2.922}_{-0.745}$ \\
	$b_1$ & $0.509(0.020)$ & ${}^{+0.020}_{-0.020}$ & & $0.511(0.100)$ & ${}^{+0.087}_{-0.108}$ & & $0.501(0.080)$ & ${}^{+0.080}_{-0.080}$ &  & $8.221(1.697)$ &${}^{+2.232}_{-0.552}$ \\
	$b_2$ & -- & -- & & -- & -- & & $0.001(0.010)$ & ${}^{+0.010}_{-0.090}$ & & $4.033(0.888)$ & ${}^{+1.171}_{-0.350}$ \\
	\enddata
	\tablecomments{The mean of different coefficients ($a_i,b_i$) with the standard deviation $\sigma$ and the $68\%$ CL. The results are obtained from  the mock SNIa data.}
\end{deluxetable}

\begin{deluxetable}{cccccccccccc}
	\tablenum{2}
	\tablecaption{\label{tab2}}
	\tablewidth{0pt}
	\tablehead{
		& \multicolumn{2}{c}{$P_{2,1}$} & & \multicolumn{2}{c}{$P_{3,1}$} & & \multicolumn{2}{c}{$P_{2,2}$} & & \multicolumn{2}{c}{$P_{3,2}$} \\
		\cline{1-3} \cline{5-6} \cline{8-9} \cline{11-12}
		\colhead{} &\colhead{\it Mean($\sigma$)} &\colhead{\it 0.68 CL} &\colhead{} &\colhead{\it Mean($\sigma$)} &\colhead{\it 0.68 CL} &\colhead{} &\colhead{\it Mean($\sigma$)} &\colhead{\it 0.68 CL} &\colhead{} &\colhead{\it Mean($\sigma$)} &\colhead{\it 0.68 CL}
	}
	\startdata
	$a_2$ & $1.381(0.218)$ & ${}^{+0.181}_{-0.242}$ & & $4.078(2.688)$ & ${}^{+1.498}_{-3.394}$ & & $1.370(0.543)$ & ${}^{+0.341}_{-0.622}$ & & $8.454(2.154)$ & ${}^{+3.113}_{-0.888}$ \\
	$a_3$ & -- & -- & & $1.241(1.265)$ & ${}^{+0.599}_{-1.520}$ & & -- & -- & & $8.834(2.474)$ & ${}^{+3.283}_{-1.607}$ \\
	$b_1$ & $0.544(0.148)$ & ${}^{+0.121}_{-0.165}$ & & $3.015(2.463)$ & ${}^{+1.313}_{-3.092}$ & & $0.532(0.443)$ & ${}^{+0.274}_{-0.509}$ &  & $7.566(2.053)$ & ${}^{+2.947}_{-0.791}$ \\
	$b_2$ & -- & -- & & -- & -- & & $0.003(0.100)$ & ${}^{+0.114}_{-0.057}$ & & $3.240(1.144)$ & ${}^{+1.577}_{-0.839}$ \\
	\enddata
	\tablecomments{ The mean of different coefficients ($a_i,b_i$) with the standard deviation $\sigma$ and the $68\%$ CL.	The results are obtained from  the Pantheon SNIa sample. }
\end{deluxetable}

\begin{figure}
	
	\gridline{
		\fig{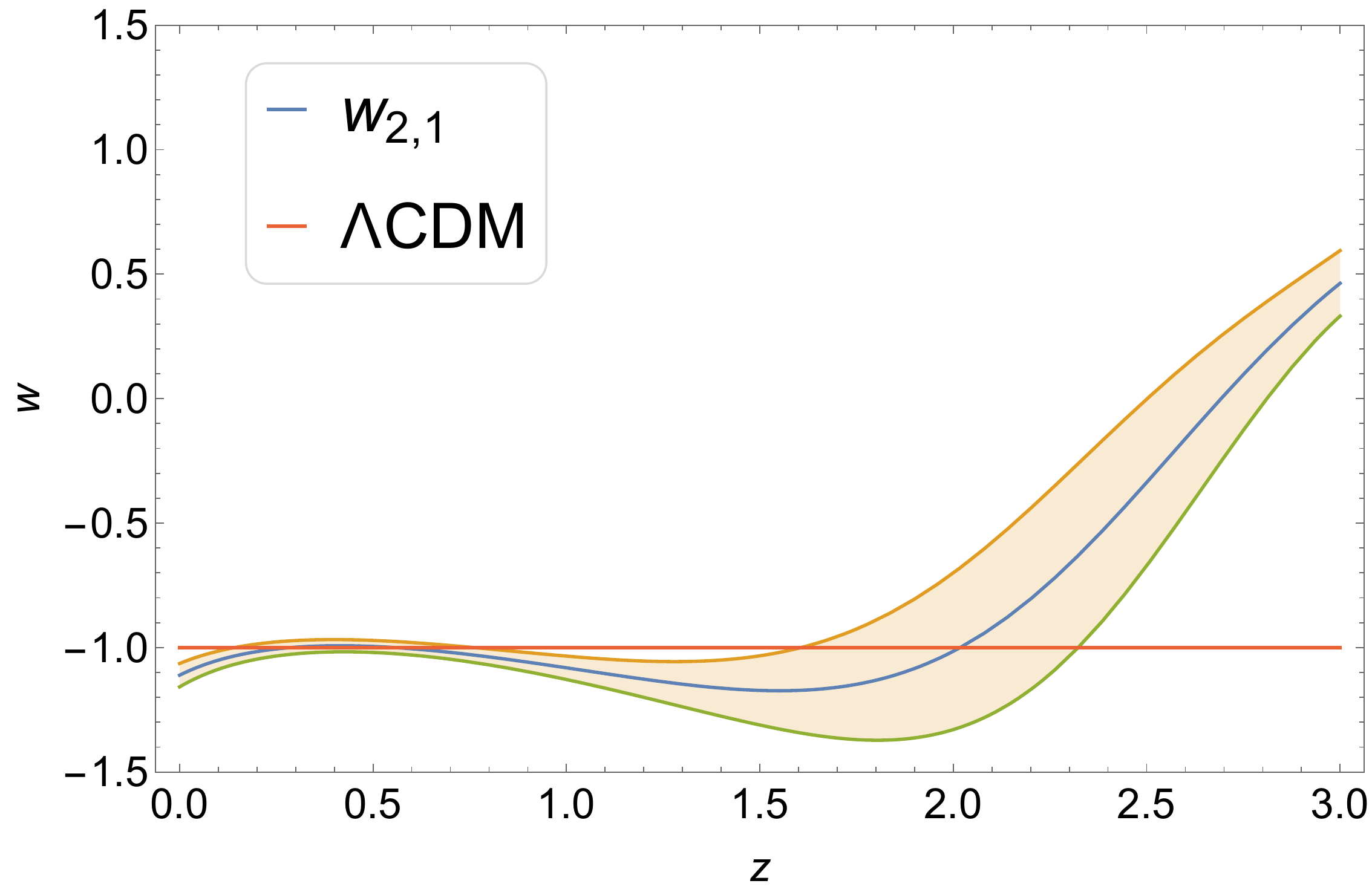}{0.5\textwidth}{(a)}
		\fig{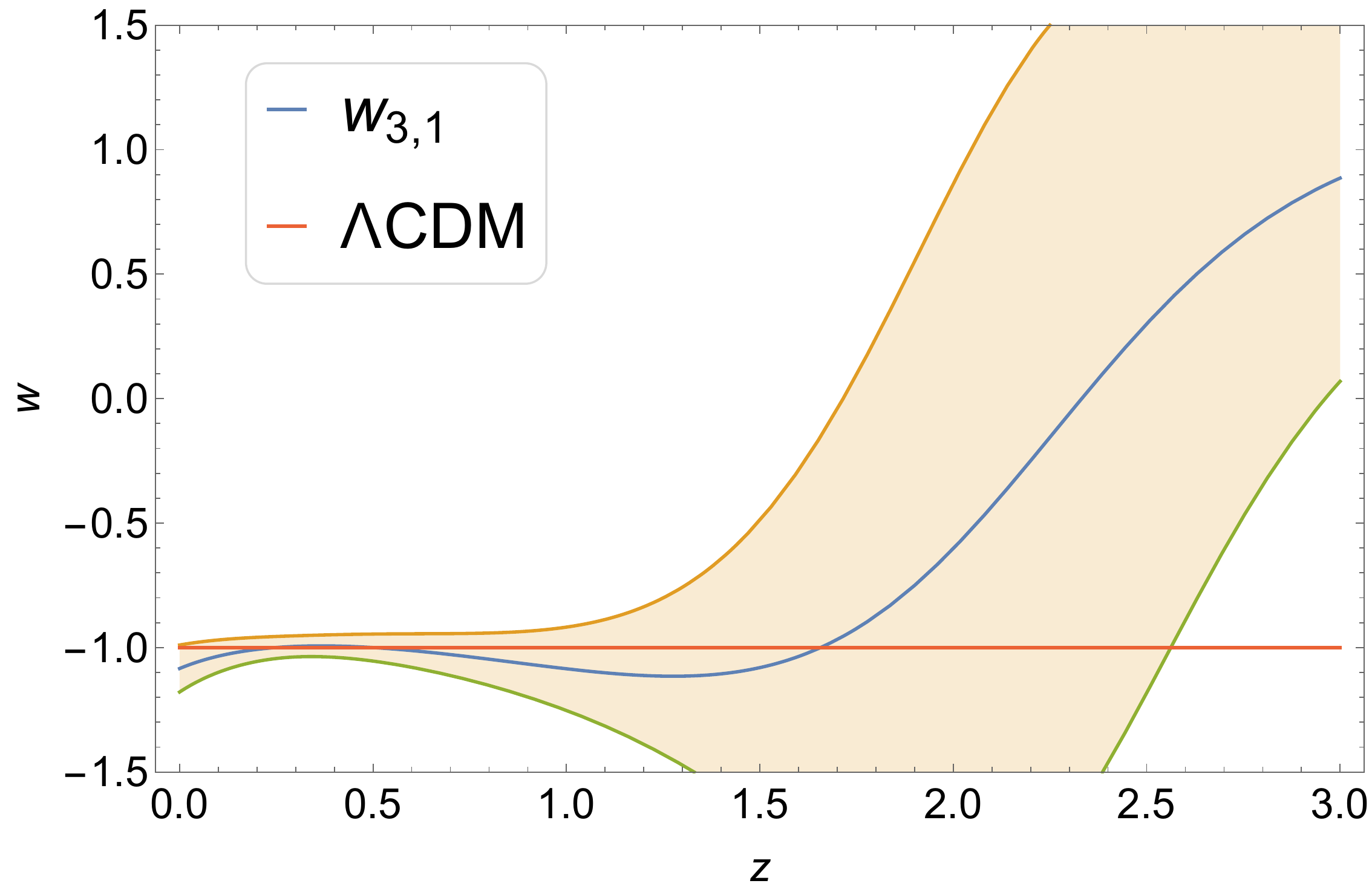}{0.5\textwidth}{(b)}
	}
	\gridline{
		\fig{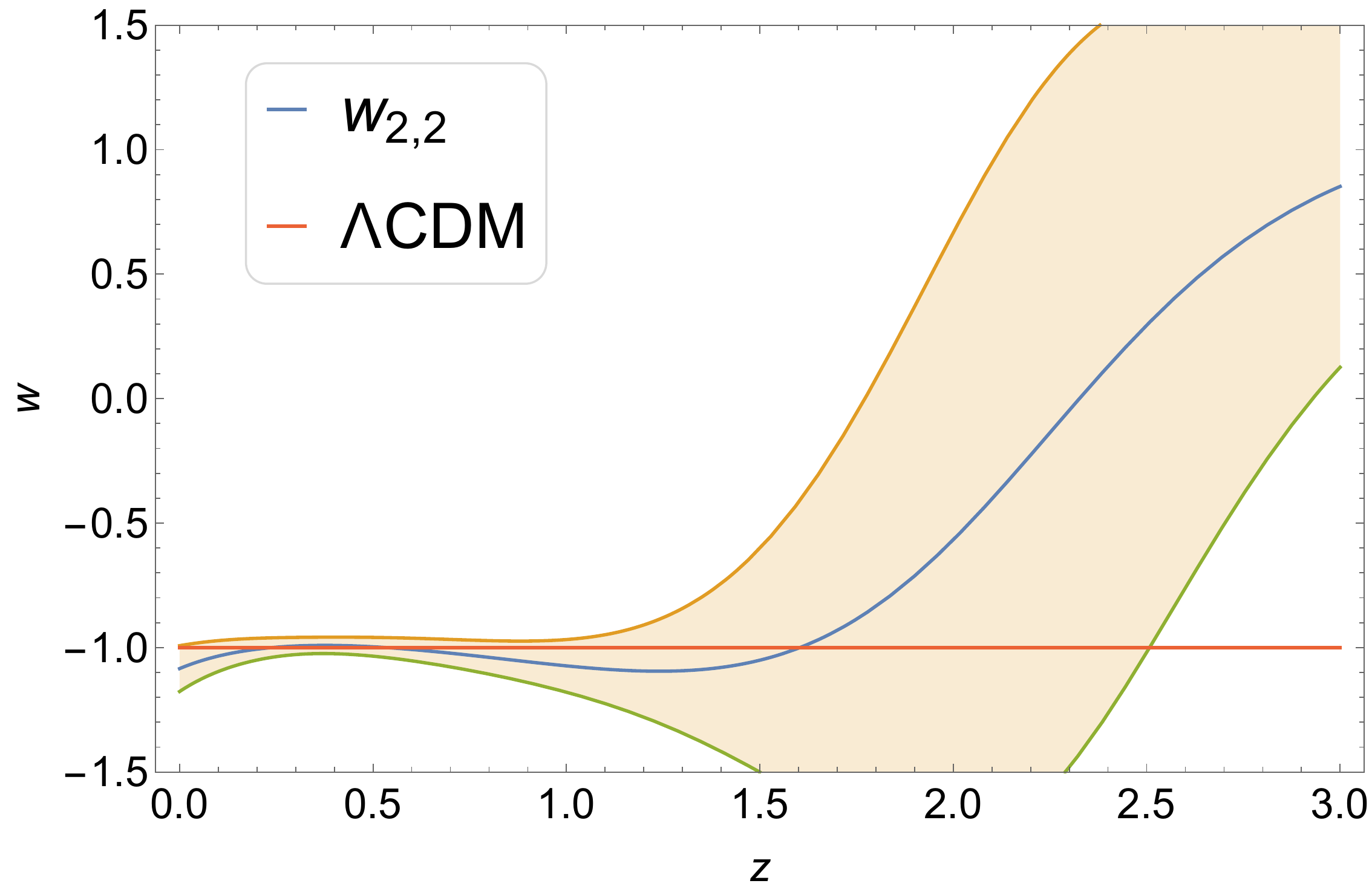}{0.5\textwidth}{(c)}
		\fig{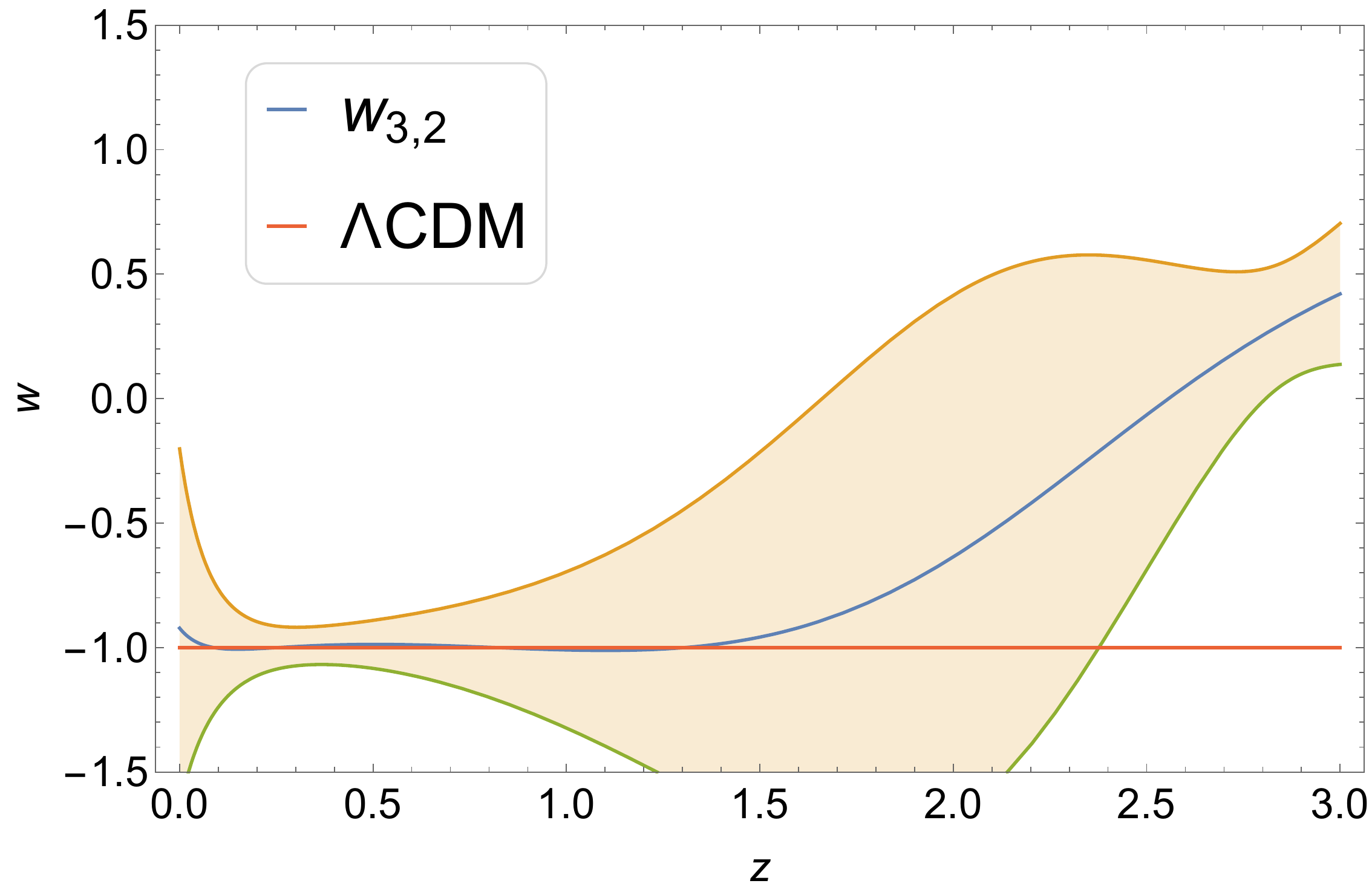}{0.5\textwidth}{(d)}
	}
	\caption{The evolution of the dark energy equation of state with $1\sigma$ error reconstructed from Pad$\acute{e}$ cosmography.  $w_{2,1}$, $w_{3,1}$, $w_{2,2}$ and $w_{3,2}$ represent the results from  $P_{2,1}$, $P_{3,1}$, $P_{2,2}$ and $P_{3,2}$, respectively.
		\label{fig4}}
	
\end{figure}

After the free parameters in the Pad$\acute{e}$ approximants having been determined by the mock data, one  can use Eq.~(\ref{w}) with $\Omega_{\mathrm m0}=0.3$ to obtain the evolution of the dark energy equation of state, which is shown in   Fig.~\ref{fig4}.  
One can see that  $w_{2,1}$ deviates from the $-1$ line in the redshift region around $1$ and in the very low redshift regions at the $1\sigma$ confidence level, while  $w_{3,1}$, $w_{2,2}$ and $w_{3,2}$ are consistent with $-1$ at the $1\sigma$ CL when $z$ is less than about $2.2$.  In the high redshift region, {\it i.e.} $z>2.8$, it is easy to see that  all of the  reconstructed $w(z)$ apparently deviate  from $-1$ line and are larger than $-1$. This is because when $z$ is very large $w_{2,1}$, $w_{2,2}$ and $w_{3,2}$  approach  $1/3$,  and $w_{3,1}$ approaches  $0$, while the dark energy  model used to mock data is the cosmological constant.

\section{conclusion}\label{conclusion}
In this paper, we investigate the reconstruction of the dark energy equation of state from the Pad$\acute{e}$ cosmography.   We first diagnose  the Pad$\acute{e}$ cosmography by using the one-parameter method and  find that it can give a description  of  the  background cosmology  better than the standard cosmography. Then,
we obtain that  the dark energy equation of state approaches a constant, {\it i.e.} $1/3$ or $0$, when the redshift is very large.  This result is general since it is independent of the coefficients in the Pad$\acute{e}$ approximant and the value of $\Omega_{\mathrm {m0}}$. This intrinsic  character will bias the $w(z)$ reconstruction 
and lead to misconception of the property of dark energy.    
By using the mock data based on the $\Lambda$CDM model, we demonstrate that the  Pad$\acute{e}$ approximant can describe the cosmic evolution very well since the relative deviation from the fiducial model is less than $0.008$. However, when  reconstructing the equation of state of dark energy,  the reconstructed results  in the high redshift regions apparently deviate from the $-1$ line, which is used to mock data. This arises as a result of the fact that the reconstructed dark energy equation of state from the Pad$\acute{e}$ cosmography approaches  a constant at very high redshifts, while the equation of state of dark energy used to mock data is $-1$.  Our results indicate that one must  exercise  caution  in reconstructing  the property of dark energy from the Pad$\acute{e}$ cosmography when the high redshift data, {\it i.e. } $z>2$, are used to constrain model parameters. Thus, the viable way to probe the property of dark energy from the Pad$\acute{e}$ approximant is to use it to directly express the dark energy equation of state~\citep{Rezaei2017,Rezaei2020} rather than the luminosity distance.

\acknowledgments
	This work was supported in part by the NSFC under Grants No. 12075084, No. 11690034, No. 11805063, and No. 11775077,  
	and by the Science and Technology Innovation Plan of Hunan province under Grant No. 2017XK2019.

\section*{Data Availability}

 Data available on request from the authors.


\end{document}